\documentclass{article}

\usepackage{arxiv}

\usepackage[utf8]{inputenc} 
\usepackage[T1]{fontenc}    
\usepackage{hyperref}       
\usepackage{url}            
\usepackage{booktabs}       
\usepackage{amsfonts}       
\usepackage{nicefrac}       
\usepackage{microtype}      
\usepackage{lipsum}

\usepackage{amssymb}
\usepackage{amsmath}

\usepackage{url}

\usepackage{hyperref}

\usepackage{float}
\input epsf
\usepackage{graphicx}
\usepackage{latexsym}

\usepackage{caption}
\usepackage{booktabs}
\usepackage{multirow}

\title{WCE Polyp Detection with Triplet based Embeddings}

\author{
  Pablo Laiz\thanks{Corresponding author: 
  E-mail address: laizpablo@ub.edu} \\
  Department of Mathematics \\
  and Computer Science, \\
  University of Barcelona, \\
  08007 Barcelona, Spain \\
   \And
  Jordi Vitri\`a \\
  Department of Mathematics \\
  and Computer Science, \\
  University of Barcelona, \\
  08007 Barcelona, Spain \\
  \AND
  Hagen Wenzek \\
  CorporateHealth \\
  International ApS\\
  \AND 
  Carolina Malagelada \\
  Digestive System Research Unit,\\
  University Hospital Vall d’Hebron, \\
  Spain
  \AND 
  Fernando Azpiroz \\
  Digestive System Research Unit,\\
  University Hospital Vall d’Hebron, \\
  Spain
   \And
  Santi Segu\'i \\
  Department of Mathematics \\
  and Computer Science, \\
  University of Barcelona, \\
  08007 Barcelona, Spain \\
}

\begin{document}
\maketitle

\begin{abstract}
Wireless capsule endoscopy is a medical procedure used to visualize the entire gastrointestinal tract and to diagnose intestinal conditions, such as polyps or bleeding. Current analyses are performed by manually inspecting nearly each one of the frames of the video, a tedious and error-prone task. Automatic image analysis methods can be used to reduce the time needed for physicians to evaluate a capsule endoscopy video. However these methods are still in a research phase.

In this paper we focus on computer-aided polyp detection in capsule endoscopy images. This is a challenging problem because of the diversity of polyp appearance, the imbalanced dataset structure and the scarcity of data. We have developed a new polyp computer-aided decision system that combines a deep convolutional neural network and metric learning. The key point of the method is the use of the Triplet Loss function with the aim of improving feature extraction from the images when having small dataset. The Triplet Loss function allows to train robust detectors by forcing images from the same category to be represented by similar embedding vectors while ensuring that images from different categories are represented by dissimilar vectors. Empirical results show a meaningful increase of AUC values compared to state-of-the-art methods.

A good performance is not the only requirement when considering the adoption of this technology to clinical practice. Trust and explainability of decisions are as important as performance. With this purpose, we also provide a method to generate visual explanations of the outcome of our polyp detector. These explanations can be used to build a physician's trust in the system and also to convey information about the inner working of the method to the designer for debugging purposes.  
\end{abstract}


\section{Introduction}

According to the Global Health Organization, colorectal cancer is the third most frequent type of cancer with 1.8 million people diagnosed in 2018 \cite{cancer_statistics}. The early detection of cancer, when it is still small and has not spread, is essential for the treatment and the survival of the patient. The detection and removal of intestinal polyps, an abnormal growth of the tissue that can evolve into cancer, is specially important. According to the American Cancer Society, screening tests of the gastrointestinal (GI) tract have significantly increased the survival rate of colorectal cancer patients\footnote[1]{https://www.cancer.org/cancer/colon-rectal-cancer/detection-diagnosis-staging/survival-rates.html}. 

The standard clinical procedure for the screening of the rectum and the colon is a colonoscopy. 
Despite the fact that this procedure is widely accepted, it has some important drawbacks: it requires qualified personnel in expensive medical facilities and may result in patient discomfort.

Wireless Capsule Endoscopy (WCE), originally developed by \cite{capsule_endoscopy}, is an alternative technique designed to visualize the inside of the digestive tract with minimal patient discomfort. 
Patients ingest a vitamin-size capsule that contains a camera and an array of LEDs powered by a battery, to record and send the captured images to an external device for a posterior analysis.

WCE has become a solution to the rapid increase in demand for optical endoscopies in recent years \cite{Li2018}, as it can deliver GI investigations without the need for expensive clinical resources and much improved patient comfort. It has been reported report that this device can accurately evaluate pathologies such as gastrointestinal bleeding \cite{USMAN201616,wce_3_dcnn_bleeding,zwinger2019capsocam}, Crohn’s disease \cite{goran2018capsule}, ulcerative colitis \cite{ozawa2019novel,maeda2019fully}, small-bowel tumors, polyposis syndrome \cite{ai_capsule_here, Yang2020} and is also applicable in polyp detection \cite{Kobaek-Larsen2018}. Furthermore, recent studies \cite{McGoran2019, Takada2020} foresee WCE as a tool to not only investigate the symptoms of GI disorders but also, in the future, to perform therapeutic interventions. The capsule could also democratize the screening process since it is better tolerated than standard endoscopy \cite{McGoran2019}, has minimal invasiveness with user-friendliness \cite{Vasilakakis2019}, does not require sedation and has fewer potential complications.

Although WCE presents many advantages over other screening techniques, it presents an important drawback in clinical practice: resulting videos can contain hundreds of thousands of images per patient that must be screened by clinical specialists. This screening is complex, tedious and time-consuming, often lasting 2 to 3 hours per video \cite{Vasilakakis2019, Yang2020}. Moreover, and also because of the fatigue caused by the visual inspection of these videos, it is common to review procedures more than once to ensure that no pathological images are missed \cite{ai_capsule_here}. All these inconveniences hinder the adoption of WCE, exposing the need of computer-aided decision (CAD) support systems \cite{aid_polyps,rahim2020survey} with artificial intelligence \cite{McGoran2019,hwang2018application,Yang2020}.

\begin{figure*}[ht]
    \centering
    \includegraphics[width=1\textwidth]{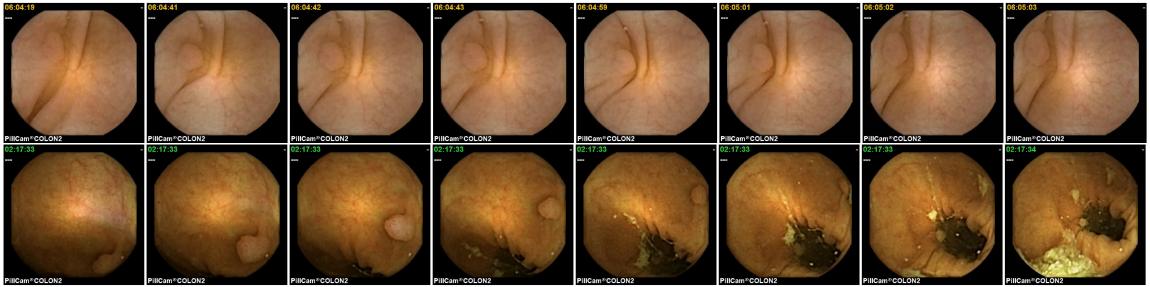}
    \caption{Illustration of two polyp sequences extracted from different patients. 
    In the first sequence can be seen how the polyp appears in all the frames approximately in the same location.
    However, in the second sequence, the polyp location change while the WCE moves through the GI tract.}
    \label{fig:sample_seq}
\end{figure*}

\begin{figure*}[ht]
    \centering
    \includegraphics[width=1\textwidth]{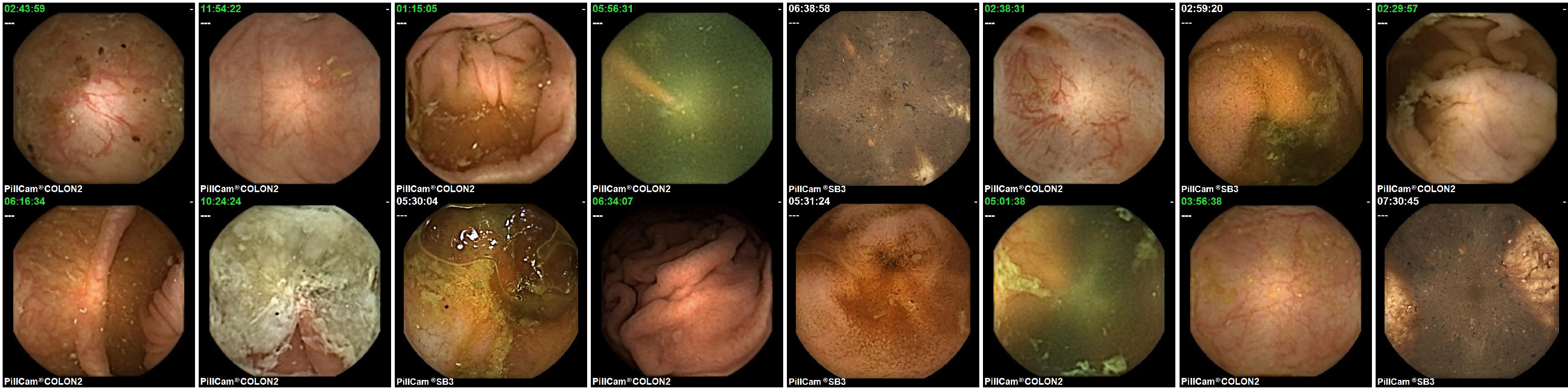}
    \caption{Illustration of 16 random samples obtained from the same procedures that represent the huge diversity of the dataset.
    For example, some of the frames present turbid, GI walls or wrinkles among others.}
    \label{fig:polyp_sample}
\end{figure*}

In the literature, we can find several AI-based CAD systems specially designed to detect suspicious or abnormal WCE images. Most of these methods are aimed at reducing visualization and diagnosis time by detecting specific GI events with high performance machine learning systems \cite{Takada2020}. 

With regard to the specific goal of polyp detection, most of the published systems have been reported and validated as automatic detection methods. However, because of legal and practical reasons, these systems cannot be used for automatic diagnosis and can only be deployed as decision support systems which filter the whole set of frames to allocate physician's attention to those images that show potential polyp structures. In most cases, this is a needle-in-haystack problem because of the occasional appearance of images with these pathologies.  
Figure \ref{fig:sample_seq} shows two sequences from different procedures where a polyp is observed. It is important to point out that, in both procedures, those are the only images of the whole procedure where a polyp is visible. 
Figure \ref{fig:polyp_sample} shows some random images from the same procedures. 

Polyp detection has been an active research topic, as it can be seen in Table \ref{tab:polyp-detection-hand-crafted}. However, to our knowledge, there is no agreement about a common evaluation methodology to allow the community to compare different CAD methods. Most of these methods have been developed and evaluated with private datasets and using different evaluation methodologies, which are suited for image detection systems but not fully informative for CAD systems in medical applications.

In this paper, we propose and validate a CAD system for the detection of polyps in WCE videos. The proposed system is based on deep Convolution Neural Networks (CNN). It is well known that CNNs have become state-of-art in many visual recognition problems, but their application in the medical field has been rather limited, with some exceptions like dermatology and breast x-rays. The main reason for this is that medical databases are comparably poor and small due to the high costs involved in data acquisition, their complex labelling, and because the use of these data often involves confidentiality issues  \cite{deepLearningChallengesMIA}.

Small size and imbalanced data are two of the main obstacles to develop reliable deep learning classifiers, because if not properly addressed, they may lead to training overfitting and poor generalization. Several tricks and techniques, such as dropout \cite{dropout}, sampling strategies \cite{sampling_strategy}, image augmentation \cite{image_augmentation} and curriculum learning \cite{Taghanaki2019}, try to alleviate this problem, but it is still an open and important challenge in the medical field as described in \cite{deepLearningChallengesMIA}. To this aim, and to overcome the small amount of available data for training the CNN, in this paper we propose an optimization strategy based on deep metric learning that uses the Triplet Loss function \cite{triplet_loss}. The obtained results show that this learning strategy outperforms the classical learning strategy using the cross-entropy loss function in our problem.

Our contributions are as follows:
\begin{itemize}
    \item We propose an evaluation methodology that involves quantitative metrics as well as the reporting of qualitative database information in order to allow fair comparisons between different systems.
    \item We show how to build an end-to-end CNN polyp detection system, based on the Triplet Loss function, that overcomes the problem of imbalanced datasets.
    \item Finally, we propose the use of classifier interpretation techniques as a mechanism to build trust in the system.
\end{itemize}

The paper is organized as follows: first, we give an overview of the related work in the field. This is followed
by a description of our methodology, presenting the system architecture, parameter optimization and evaluation methodology, followed by experimentally setup and results. Finally, we conclude the paper and give directions for future work.

\section{Related Work}

Since the presentation of WCE, several computational systems have been proposed to reduce its inherent drawbacks in clinical settings \cite{liedlgruber2011computer, belle2013biomedical}. Generally, these systems are designed either for efficient video visualization \cite{mackiewicz2008wireless, chu2010epitomized, iakovidis2013efficient, drozdzal2013adaptable} or to automatically detect different intestinal abnormalities such as bleeding \cite{wce_3_dcnn_bleeding, USMAN201616, wce_4_haemorrhage}, polyp \cite{rotation_invariant, ploosonePolyp}, tumor \cite{cobrin2006tumor}, ulcer \cite{Ciaccio2013ulcer}, motility disorders \cite{Malagelada2015motility, sseguiWrinkles} and other general pathologies \cite{ciaccio2010distinguishing, 6051474, malagelada2012functional, Chen2013, ZHAO2015108}. Deep learning nowadays represents the state-of-the-art to most of these problems. Table \ref{tab:wce-deep} shows detailed information of those systems that have been implemented using deep learning methods.

\begin{table*}[ht]
    \centering
    \caption{Comparison of existing Deep Learning methods for the classification problem in WCE.  In the last column, Metrics, the legend used is: Accuracy (A), Sensitivity - Recall - TPR (B), Specificity - TNR (C), ROC (D), AUC (E), Precision (F), Confusion Matrix (G), F1-Score (H), Cohen's Kappa score (I).}
    \label{tab:wce-deep}
    \resizebox{\linewidth}{!}{%
        \begin{tabular}{cccccccc}
        \toprule
        \textbf{Reference (Year)} & \textbf{Class} & \multicolumn{2}{c}{\textbf{Dataset}} &  \multicolumn{2}{c}{\textbf{Validation}} & \textbf{Architecture} & \textbf{Metrics} \\ 
         &  & Videos & Images & Method & Patient Separation &   \\
        \midrule
        \cite{8_classifying_digestive_organs}  & Localization & 25    & 75k   & 60k / 15k & Unknown & AlexNet & A\\ 
        \cite{9_hybrud_conv}  & Digestive organs & 25 & 1M    & 60k / 15k & Unknown & CNN + ELM & A\\ 
        \cite{wce_2_generic_feature}  & Scene classification  & 50    & 120k  & 100k / 20k & Unknown & CNN & A-G\\ 
        \cite{wce_3_dcnn_bleeding}  & Bleeding & - & 10k   & 8.2k / 1.8k & Unknown & AlexNet & B-F-H\\ 
        \cite{wce_4_haemorrhage}  & Haemorrhage & - & 11.9k & 9.6k / 2.24k & Unknown & LeNet, AlexNet, & F-B-C-H\\ &&&&&& GoogleNet, VGG-Net \\ 
        \midrule
        \end{tabular}
    }
\end{table*}

Among possible WCE uses, polyp detection has been one of the problems that have attracted a lot of attention from researchers. Table \ref{tab:polyp-detection-hand-crafted} presents a set of methods, published in high impact conferences and journals, aimed at detecting polyps by using any of the GI examination modalities. As it can be seen, prior to 2015, most of the published methods were based on conventional computer vision and machine learning techniques, which are based on the extraction of handcrafted visual features followed by a classifier. These systems have used several image features such as color, texture and shape to deal with the classification task.

\begin{table*}[ht]
    \centering
    \caption{Overview of our proposed and existing method for polyp detection. The nomenclature is the same as in Table \ref{tab:wce-deep}.}
    \label{tab:polyp-detection-hand-crafted}
    \resizebox{\linewidth}{!}{%
        \begin{tabular}{ccccccccc}
        \toprule
        \textbf{Reference (Year)} & \textbf{Modality} & \multicolumn{3}{c}{\textbf{Dataset}} & \multicolumn{2}{c}{\textbf{Validation}} & \textbf{Feature} & \textbf{Metrics} \\
         &  & Videos & Polyp & Non-polyp & Method & Patient Separation &  \\
        \midrule
        \cite{2_intestinal_polyp_recognition}  & WCE & 2  & 150 & 150 & 3-fold & Unknown & Colour and shape  & A-B-C\\ 
        \cite{1_poylp_detection_color_texture_features}   & WCE & 2  & - & -  & 5-fold & Unknown & Colour  & A-D-E\\ 
        \cite{5_automatic_polyp_detection}   & WCE & 10 &  600 & 600    & 10-fold & Unknown & Texture & A \\ 
        \cite{3_feature_polyp_detection}  & WCE & 10 & 436  & 436  & 10 random splits & Unknown & Texture  & A-B-C\\ 
        \cite{7_polyp_detection_imbalanced}  & Endoscopy & 141 & 1k & 100k & 5-fold & Unknown & Shape  & B-E-F\\ 
        \cite{4_automatic_polyp_detection}  & Colonoscopy & 20 & 2k & 3k & - & Unknown &Texture & B\\ 
        \cite{10_lesion_dtection}  & Endoscopy  & - & 6.5k & 50k  & 10-fold & Unknown & CNN  & A-B-C\\ 
        \cite{endo_3_automatic_detection}  & Colonoscopy & - & 826 & 1.1k   & Random test & Unknown & CNN   & A-C-F-H\\ 
        \cite{endo_4_automatic_polyp_detection}   & Colonoscopy   & 6 & 37k &  36k   & Random test & Unknown & CNN & A-B-G\\ 
        \cite{endo_5_integrating_online}  & Colonoscopy & 20 & 3.7k &  - & 18 dif. videos & Separate &  FCN & G-F-B-H \\ 
        \cite{11_deep_polyp_recognition}  & WCE & 35 & - & -  & 1k /3k & Unknown & SAE &  A-G\\ 
        \cite{rotation_invariant}  & WCE & 62 & 1.5k & 1.5k &  150/150 & Unknown & CNN & A-B-F-H\\ 
        \cite{ploosonePolyp}  & Colonoscopy & 215 & 404 & - & 50 rand. images & Unknown & CNN & G-F-B-H \\ 
        \cite{Yuan2019DenselyCN}  & WCE & 80 & 1.2k & 6k  & 120/600 & Unknown & CNN  &  A-B-C-F-H\\ 
        \cite{Guo2019TripleANet}  & WCE & - & 585 & 2.2k  & 4-fold & Unknown & CNN &  A-I\\ 
        This Paper & WCE & 120 & 2.1k & 1.3M &  5-fold & Separate  & CNN  & \\
        \bottomrule
        \end{tabular}
    }
\end{table*}

Since 2015, and following the success of deep learning methods in any computer vision application, most of the proposed methods to detect WCE events are based on deep learning. One of the first methods, know as SSAEIM, was proposed by \cite{11_deep_polyp_recognition}. This method, which uses a set of concatenated sparse autoencoders and a reconstruction loss to automatically find powerful features for the classification task, shows an improvement over previous methods based on handcrafted features.

\cite{rotation_invariant}, 
argued that object rotation and high intra-class variability are two main limitations for WCE image analysis. In order to overcome this problem, the authors proposed a new method named RIIS-DenseNet, based on a DenseNet, which uses two loss functions as constraints. The rotation-invariant constraint was designed to achieve rotation invariance by enforcing similarity between feature representations of the training samples and their corresponding rotated ones. Meanwhile, the image similarity constraint was proposed to allow a small intra-class variance in the feature space.

The same authors \cite{Yuan2019DenselyCN} proposed DenseNet-UDCS one year later, aiming to overcome three different problems: imbalanced database, small inter-class variance and high intra-class variability. To achieve these goals, the network uses a weighted cross-entropy loss together with a category sensitivity loss. Weighted cross-entropy is appropriate to deal with the imbalanced dataset while category sensitive loss aims to reduce the distances between feature representation of samples from the same category.

\cite{Guo2019TripleANet} proposed Triple ANet, a CNN system which addresses the problems of high intra-class variability, small inter-class variance and the existence of artefacts in the images. The system introduces two blocks of deformable convolutions to capture correlations and highlight informative areas in the images. The other fundamental point of the system is the replacement of the classifier by an angular contractive loss. 

These methods represent a significative achievement, overcoming some of the main problems related to WCE image analysis, but we think that the solution of this problem is not complete if the right metrics and evaluation methodologies are not used for validating them.  There  are  three  features  of  these  methods  that  are  worth to analyze in order to define right comparison methodology: database size, validation strategy and evaluation metrics.

\textbf{Databases:} As it can be seen, in most of the cases the number of polyps in the dataset is relatively small. The paper that uses the largest number of polyps uses a total of 37,000 images, while the smallest uses just 25. If we consider only those papers which work exclusively with WCE images versus also colonoscopy images, the number of images is significantly smaller. The paper that uses the largest dataset uses a total of 1,500 polyp images obtained from 62 different patients, which means an average of 25 polyp frames per procedure. It is also important to point out that the number of procedures is 2 to 1,000 times smaller than the number of polyps. This means that several images from the same patient or even polyp are used in the dataset, but this information is not usually reported. Besides this and in order to understand how challenging the dataset is, it would also be important to report the size and type of polyps. 
Regarding negative samples, the paper that uses the largest databases uses a total of 100,000 images while the paper that uses the smallest set uses 75 images. No information about the sampling strategy that was used to obtain these negative images is reported in any case.

\textbf{Training and validation strategy:} As pointed out before, datasets usually contain several positive images from the same patient, and in most cases several images from the same polyp. For this reason, it is very important to ensure that the training and the validation sets do not share images from the same procedure. If the partition of the training and validation is not properly done, it would be highly probable to have consecutive and practically identical frames in both sets. This fact clearly contaminates any validation result based on those datasets. Only the method presented by \cite{endo_5_integrating_online} reports this information.

\textbf{Evaluation metrics:} In order to validate these systems, authors use a variety of evaluation metrics: accuracy, precision, sensitivity/recall, specificity, ROC-Curve, AUC, F1-Score as well as the confusion matrix. The diversity of evaluation metrics clearly hinders a clear comparison between methods, thus showing the need for a good and unified evaluation strategy.

\section{Method}
Taking into account that we are dealing with a computer-aided decision system, 
we designed and evaluated our approach not as a classical classification problem, but as an information retrieval problem. Given a WCE video, the problem is to rank the images of the video according to some criterion so that the more relevant images appear early in the result list displayed to the physician. This allows the visual screening of a reduced set of images and at the same time ensures the detection of a maximum number of positive images.

The description of the method is divided in the following three parts:
\begin{itemize}
    \item System Architecture: Introduction of the CNN architecture used to detect polyp images. 
    \item Parameter Optimization: Explanation of how the chosen architecture is optimized. Presentation of the problems derived from the database and how to adapt the learning process to achieve better results. 
    \item Evaluation Methodology: Presentation of standard metrics and discussion about how to evaluate polyp detection systems to be able to compare them with other works.
\end{itemize}

\subsection{System Architecture}

The proposed deep learning method is based on the Deep Residual Network (ResNet) architecture, presented by \cite{resnet_paper}. This network has shown outstanding results in important image recognition problems. 

The main novelty of this architecture is the use of a high number of layers that progressively allow to learn more complex features. The first layers learn edges, shapes or colors while the last ones are able to learn concepts.
In order to learn, this architecture needs the introduction of a set of {\em residual blocks} that avoid the problem of vanishing gradients when the number of layers increases. These blocks are built by using skip connections or identity shortcuts, that reuse the outputs from previous layers. 

The residual block has the following form:
\begin{equation}
    y = F(x, \{W_i\}) + x
\end{equation}
where $F(x, \{W_i\})$ represents stacked non-linear layers and $x$ the identity function.

Taking into account the performance of this architecture in other image classification problems, we used the fifty-layers ResNet version, known as ResNet-50.

\begin{figure*}[ht]
    \centering
    \includegraphics[width=1\textwidth]{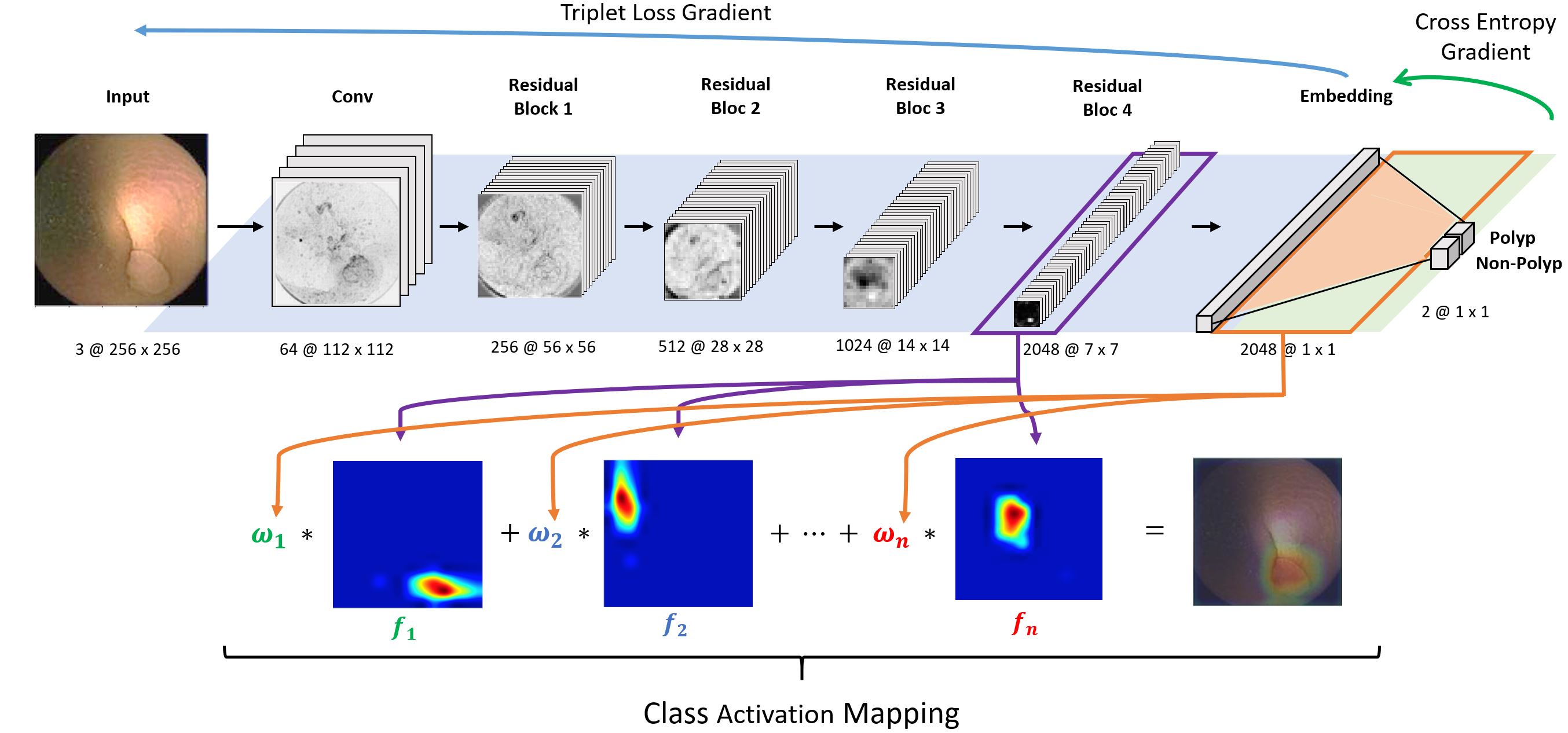}
    \caption{
    Overview of the proposed CNN structure.
    The upper part of the scheme appear the ResNet architecture with our methodology applied in it. The background colour reflects the layers affected by each one of the gradient generated by the main losses.
    The lower part of the figure shows how the class activation mapping is built. 
    }
    \label{fig:architecture}
\end{figure*}

\subsection{Parameter optimization}

ResNet-50 has over 23 million trainable parameters. The robust estimation of these parameters needs of millions of images as described in \cite{resnet_paper}, but these parameters have been shown useful for a variety of visual recognition problems. The original paper used the cross-entropy loss function with a L2 regularization term  to estimate all these parameters. Binary cross-entropy loss function decreases as the prediction converges to the true label, and increases otherwise, as its function indicates:
\begin{equation}
    L_{CE}(p, y) = - y \cdot \log(p) - (1-y) \cdot \log(1-p)
\end{equation}
where $y$ is the true label of the sample and $p$ is the estimated probability of the sample to belong to class $1$.

In our case, to deal with a small and imbalanced dataset, we propose an optimization of the ResNet in two stages. First, images are projected into an embedding space using the Triplet Loss (TL) as described in \cite{triplet_loss}. Then, we consider the cross-entropy loss function in the embedding space. The proposed methodology is shown in the upper part of Figure \ref{fig:architecture}.

TL, a deep metric learning (ML) method, has shown great generalization results when dealing with a large amount of classes as for instance in the problem of face identification. The goal of TL is to optimize a latent feature space $f(x) \in \mathbb{R}^d$ such that examples from the same class are closer to each other than to those belonging to different classes.

\begin{figure}[!ht]
    \centering
    \includegraphics[width=.35\textwidth]{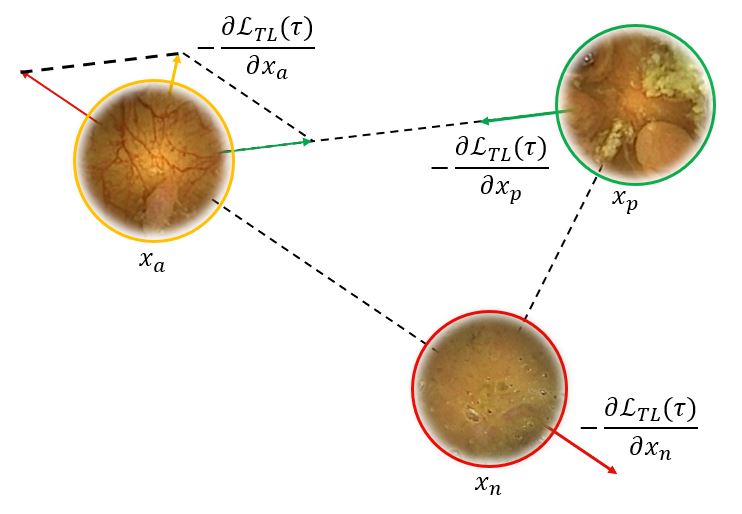}
    \caption{Behaviour representation of the of TL using one triplet. The arrows of each image indicate the direction in which the embedding will move following the gradient. }
    \label{fig:triplet_loss}
\end{figure}

In order to learn this embedding representation, TL aims at ensuring that an anchor image, $x_a$, is closer to all other images from the same class, $x_p$, than any image from a different class, $x_n$. This concept, illustrated in Figure \ref{fig:triplet_loss}, can be formulated as follows:
\begin{equation}
    \left\|{f^a-f^p}\right\|^2_2 + \alpha < \left\|{f^a-f^n}\right\|^2_2,  \forall (x_a, x_p, x_n) \in \tau
    \label{eq:triplet_idea}
\end{equation}
where $f^k$ is the embedding of $f(x_k)$, $\left\|{\cdot}\right\|_2$ is the Euclidean distance and $\alpha$ is a margin, which define the minimum distance between elements of different classes.

In order to train the network and reach the sought condition, the TL function is defined as follows:
\begin{equation}
    L_{triplet-loss} = \sum_{i=1}^{N} \Big[ \left\|{f_i^a-f_i^p}\right\|^2_2 - \left\|{f_i^a-f_i^n}\right\|^2_2 + \alpha \Big]_{+}
    \label{eq:triplet_loss}
\end{equation}

Training a neural network using TL is not simple. At training time, the network receives triplets of images. For small datasets, the generation of each triplet is feasible, but when the amount of images increases, the number of possible triplets grows with cubic complexity. If we try to generate all of them, it becomes intractable and inefficient, making it impossible to optimize the loss.
As a consequence, a sampling strategy for the images becomes an essential part of the learning method. The right choice of triplets can increase the speed of convergence, the probability to find a lower minimum and even the possibility of getting better generalization.

In the literature we can find two different methodologies to face the problem of triplet sampling for each batch.
The first one is called \textit{Batch All}, being introduced in \cite{batch_methodologies}, $TL_{BA}$. In this case, for each sample $x_a$ in the batch, we consider all possible triplets. This results in $k_0 \cdot k_1 \cdot (k_0 + k_1 -2)$ elements.
The $TL_{BA}$ loss function is:
\begin{equation}
    L_{BA}(\tau) = \sum_{c=1}^2 \sum_{a=1}^{k_0} \sum_{\substack{p=1 \\ p \neq a}}^{k_0} \sum_{n=1}^{k_1}
    \Big[ \left\|{f^a-f^p}\right\|^2_2 - \left\|{f^a-f^n}\right\|^2_2 + \alpha \Big]_{+}
\end{equation}

The use of the previous methodology declined from the appearance of the \textit{Batch Hard} \cite{batch_methodologies} approach, $TL_{BH}$. It takes each anchor $x_a$ and generate triplets by seeking in the batch for the hardest positive sample $x_p$, defined as farthest positive sample $x_p = argmax_{x_i}(\left\|{f_i^a-f_i^p}\right\|^2_2)$, and the hardest negative sample $x_n$, defined as the closest negative sample $x_n = argmin_{x_i}(\left\|{f_i^a-f_i^n}\right\|^2_2)$.
The $TL_{BH}$ loss function is:

\begin{equation}
    \begin{split}
        L_{BH}(\tau) = \sum_{a=1}^{k_0 + k_1} 
        \Big[ \underset{x^p}{argmax}(\left\|{f^a-f^p}\right\|^2_2) - \\
        \underset{x^n}{argmin}(\left\|{f^a-f^n}\right\|^2_2) + \alpha \Big]_{+}
    \end{split}
    \label{eq:batch-hard}
\end{equation}

\subsection{Evaluation}
In the field of the medical imaging, and in particular when databases are protected and not released to the public domain, the evaluation of different proposals is perhaps the hardest and most critical part. 
However, and as shown in the related work, a unified procedure which allows a objective comparison of methods does not exist. We can see that a diversity of metrics are used and in most of the cases, not the ideal ones for the problem. Moreover, in most of the papers the used or the reported information about the dataset is not sufficient to understand the relevance of the proposal. 
To this aim, in this section we study and propose a methodology to be used in order to validate computer-aided polyp detection systems.
The proposed evaluation methodology is divided into three fundamental points:
\begin{itemize}
    \item Databases and cross-validation strategy: How to properly build it and what information must be reported to understand the relevance of the proposal and allow a detailed comparison of models.
    
    \item Quantitative Results: Standard metrics in computer vision problems have several drawbacks that can affect the understating of the performance of the methods. For this reason we propose and justify a set of metric to be used.
    
    \item Qualitative Results: Aside the numeric results, it is important to consider qualitative results to trust in the system. To this aim, we propose the use of a method to understand the output of the model. 
\end{itemize}

\subsubsection{Databases and cross-validation strategy}

The creation of medical databases is an essential step before training and validating any type of system. Both, positive and negative samples must be collected in the best way. With respect to size and diversity, training data can follow any distribution that one deems appropriate, however, it is crucial that results are reported using a test set large enough to also capture the diversity of non-pathological images that can be found. In order to capture this diversity, we consider a uniform time sampling strategy as the best option for creating the negative set. As negative samples are cheap, since we have as many as needed, a minimum number of images per videos should be used, being 2,000 a reasonable number.

\begin{figure}[H]
    \centering
    \includegraphics[width=.48\textwidth]{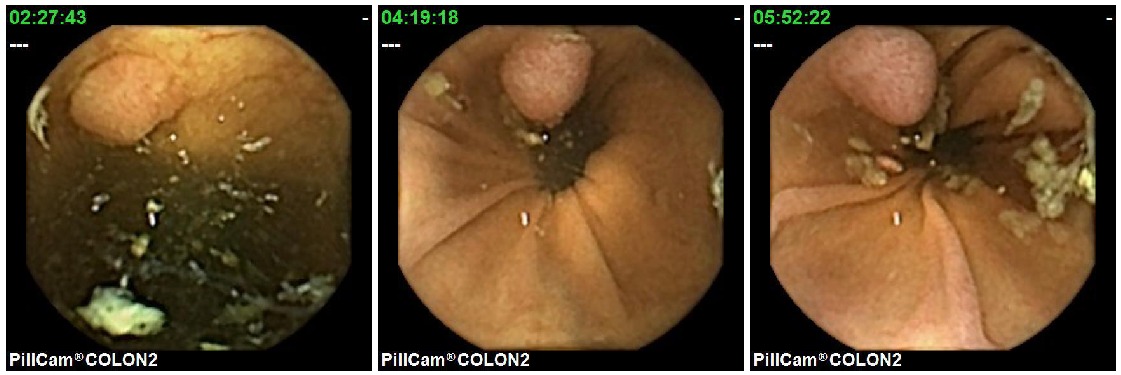}
    \includegraphics[width=.48\textwidth]{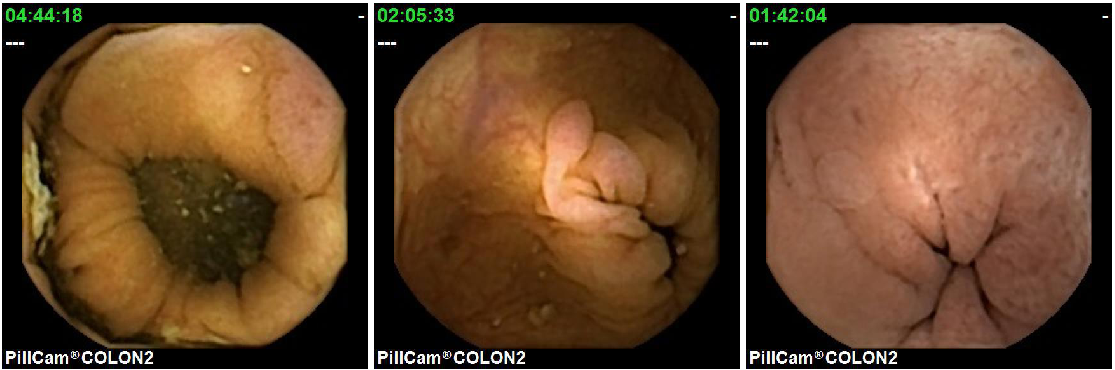}
    \caption{Example of polyps extracted from the procedures. In the first row, the polyps come from the same procedures, while the polyps from the second row come from different ones.}
    \label{fig:polyps_samples}
\end{figure}

The second important point to consider when creating the database and its evaluation methodology, is that although all polyps have common visual characteristics, the appearance of different polyps from the same patient must be considered. The first row of Figure \ref{fig:polyps_samples} shows three different polyps from the same patient, while the second row shows three polyps from different patients. As it can be observed, those polyps from the same patient are generally similar in shape and texture while the polyps from other patients are more diverse. It is for this reason, that training and test set must not use images from the same procedures.

Additionally, since the datasets are small, it is recommended to perform cross-validation to avoid data selection problems. As mentioned before, it is important that the folds of the the cross-validation process are done by leaving procedures out, not by leaving images out, therefore ensuring that images from the same procedure never belong to two different partitions.

Lastly, and since in most of the cases databases are not released to the public domain, it is fundamental to have a detailed description of the dataset in order to understand the complexity and impact of the solution. We consider that the following information should necessarily be reported:
\begin{itemize}
    \item Number of procedures/cases used in the dataset.
    \item How many of them suffer a pathology.
    \item Distribution of unique pathological events.
    \item Frames per each pathological event.
\end{itemize}

\subsubsection{Quantitative Results}

Evaluation metrics illustrate the performance of the system and allow to compare.
For this reason, they require a high capability of discrimination among models and they must match with the aim of the system.

Accuracy is the most frequently used metric to validate polyp detection systems. However, it does not reflects what is expected from the system since it does not necessarily correlate with the time needed to reach a diagnosis by the physician. Accuracy depends on the defined threshold of the system, without giving the full picture of system output. Moreover, in imbalanced problems, accuracy is mostly affected by the predominant class. Weighted accuracy is a more suitable metric, although it is still dependent on a fixed threshold.

Precision and recall (also known as sensitivity), have also been widely used by the community. Precision is the fraction of true polyp images (TP) among all the positives obtained by the system (TP + FP), while recall is the fraction of true polyp images (TP) that have been detected from the total amount of polyp images (TP + FN). However, with accuracy, these measurements are also affected by the threshold of the classifier. Since the goal of the system is to reduce the time needed for the physician, it is interesting to report the obtained recall scores at different specificity (TNR) values instead of using the best trade-off between both metrics. The recall at these fixed specificity values allows to understand the expected amount of images that are needed to be visualized by the physician in order to obtain certain performance, i.e., recall at the specificity of 95\% measures the percentage of detected polyps (TPR) if only 5\% of the images are reviewed. The recall for specificity values of $80\%$, $90\%$ and $95\%$ are analyzed for this paper.

The Area Under the Curve (AUC) is another interesting measurement. The AUC is computed from the ROC curve which relates the specificity and the recall obtained for all the possible thresholds of the classifier.
The AUC value can be understood as the probability of the classifier to predict a true positive element as a positive with higher probability than as a negative; therefore the larger the AUC value is, the better the classifier is. A limitation of the AUC is that both negative and positive classes have the same impact on the output, so FP and FN penalize equally.

\subsubsection{Qualitative Results} 

Although deep learning has shown impressive results, its application to the medical filed carries worries and criticisms since computerized diagnostic methods are seen as black boxes which do not show how the data is analyzed or how the output is obtained. In medical imaging problems, and particularly on the topic of polyp detection, it is transcendent to trust and understand the obtained predictions by the system. Understanding how the outcome was obtained allows to: 1) understand why something is considered pathological; 2) provide the needed trust of physicians and scientists in the system; 3) debug and identify errors of the system in an easier way.

To this aim, we consider that a qualitative evaluation, showing where and why the system is failing is a very important element. It is not the same to fail in a small and or partly-occluded polyp than missing a large polyp. It is also important to show False Positive (FP) cases, since these images with shapes or textures that are similar to polyps may be understandable errors and increase the confidence in the system. 

To study where the system detected a polyp in a frame can be useful for two main reasons: 1) identify if the detector is focusing in the area of interest and 2) help physicians in the review process.

Class activation map (CAM) presented by \cite{cam}, is a generic localized deep representation that could be used to interpret the prediction decision made by the system.
This method indicates the implicit attention that the network gives to the pixels of an image considering the class where it belongs. 

To obtain the class activation map a linear combination is computed between the feature maps and the classifier weights, since they connect the output and the last feature maps, which identify the importance of each response obtained.
For a given class $c$, the formalization of the class activation map $M_c$ is defined as:
\begin{equation}
    M_c(x,y) = \sum_{k} w^{c}_{k} \cdot f_{k}(x,y)
\end{equation}
where $K$ is the number of channels, $f_k$ are the different responses of the filters and $w^{c}_{k}$ is the weight that relates class $c$ with filter $k$, which is activated by some visual pattern within its receptive field.
Briefly, this technique is a weighted linear sum of these visual patterns at different spatial locations, which give the most relevant regions for the system.

\subsection{Guidelines}

After analyzing and discussing each one of the previous aspects, in order to get a full validation of the system, we propose that the following items should be included in the validation methodology:

\begin{enumerate}
    \item A fully detailed report of the dataset used.
    \item Training and validation set must not contain images from the same procedure.
    \item The negative images in the validation set must represent the diversity of the domain. A random sampling or uniform time sampling from the same videos are good strategies (patients and control cases).
    \item The number of negative images in the validation set must be higher than the number of positive images. At least 2,000 times the number of positive images should be considered.
    \item For small datasets it is necessary to apply a cross-validation method.
    \item Recall@80, Recall@90 and Recall@95 should be used in order to make the system comparable with other methods in the community.
    \item A qualitative evaluation is recommended to build trust in the system.
\end{enumerate}

\section{Experimental Setup and Results}

\subsection{Dataset Details}
The database used in this paper is composed of $120$ procedures from different patients. All these procedures have been performed using Medtronic PillCam COLON2 or PillCam SB3. 
Each image from the video was labeled as positive, where at least one polyp was visible, or negative. All these labels were obtained by expert physicians and trained nurses. Each video was examined by at least two experts. In case of controversy between experts, the final decision was taken by a final expert.  Polyps were found in $52$ out of the $120$ analyzed procedures. From those $52$ procedures with polyps, a total of $165$ different polyps were annotated and used as a positive set. Table \ref{tab:unique_polyps} summarizes the amount polyps found per procedure. As it can be observed, the number of polyps per procedure is diverse, in a majority of procedures, the experts have not reported any polyp, being $1.37$ the average of reported polyps per procedure and $11$ the maximum number in a single procedure. Table \ref{tab:frames_per_polyp} shows the number of frames where each polyp is visualized within the video. Since most polyps are observed in more than one frame, a total $2,136$ images with polyps  have been considered as positive images. 
Additional details of the database are reported in Table \ref{tab:morphology-size}, such as the morphology and size of the polyps. The size of the polyps was determined using the Rapid PillCam Software V9.

\begin{table}[H]
    \centering
    \caption{ Amount of polyps per procedure.}
    \label{tab:unique_polyps}
    \resizebox{!}{!}{
        \begin{tabular}{@{}c|ccccccccc@{}}
        \toprule
         \textit{\textbf{\# Polyps}} & \textbf{0} & \textbf{1} & \textbf{2} & \textbf{3} & \textbf{4} & \textbf{5} & \textbf{6} & \textbf{7} & \textbf{11} \\ \midrule
        \textit{\textbf{\# Procedures}} & 68 & 17 & 11 & 8 & 3 & 5 & 3 & 2 & 3 \\ \bottomrule
        \end{tabular}
    }
\end{table}

\begin{table}[H]
    \centering
    \caption{ Amount of frames per polyp.}
    \label{tab:frames_per_polyp}
    \resizebox{!}{!}{
        \begin{tabular}{@{}c|cccccc@{}}
        \toprule
        \textit{\textbf{\# Frames}} & \textbf{1-2} & \textbf{3-4} & \textbf{5-6} & \textbf{7-10} & \textbf{11-20} &
        \textbf{21+} \\ \midrule
        \textit{\textbf{\# Polyps}} & 33 & 32 & 20 & 19 & 31 & 30\\ \bottomrule
        \end{tabular}
    }
\end{table}

\begin{table}[H]
    \centering
    \caption{ Morphology - Size of the polyps}
    \label{tab:morphology-size}
    \resizebox{!}{!}{
        \begin{tabular}{ccccc|c}
        \bottomrule
         &  & \multicolumn{3}{c}{\textbf{Morphology}} &  \\ \cline{3-6} 
         &  & \textbf{Sessile} & \textbf{Pedunculated} & \textbf{Undefined} & \textbf{Total} \\ \hline
        \multirow{5}{*}{\textbf{Size}} & \begin{tabular}[c]{@{}c@{}}\textbf{Small} \\ (2-6 mm)\end{tabular} & 65 & 4 & 19 & 88 \\
         & \begin{tabular}[c]{@{}c@{}}\textbf{Medium}\\ (7-11 mm)\end{tabular} & 29 & 4 & 20 & 53 \\
         & \begin{tabular}[c]{@{}c@{}}\textbf{Large}\\ (12+ mm)\end{tabular} & 8 & 3 & 13 & 24 \\ \hline
         & \textbf{Total} & 102 & 11 & 52 & 165\\
         \bottomrule
        \end{tabular}
    }
\end{table}

\begin{figure}[h]
    \centering
    \includegraphics[width=.48\textwidth]{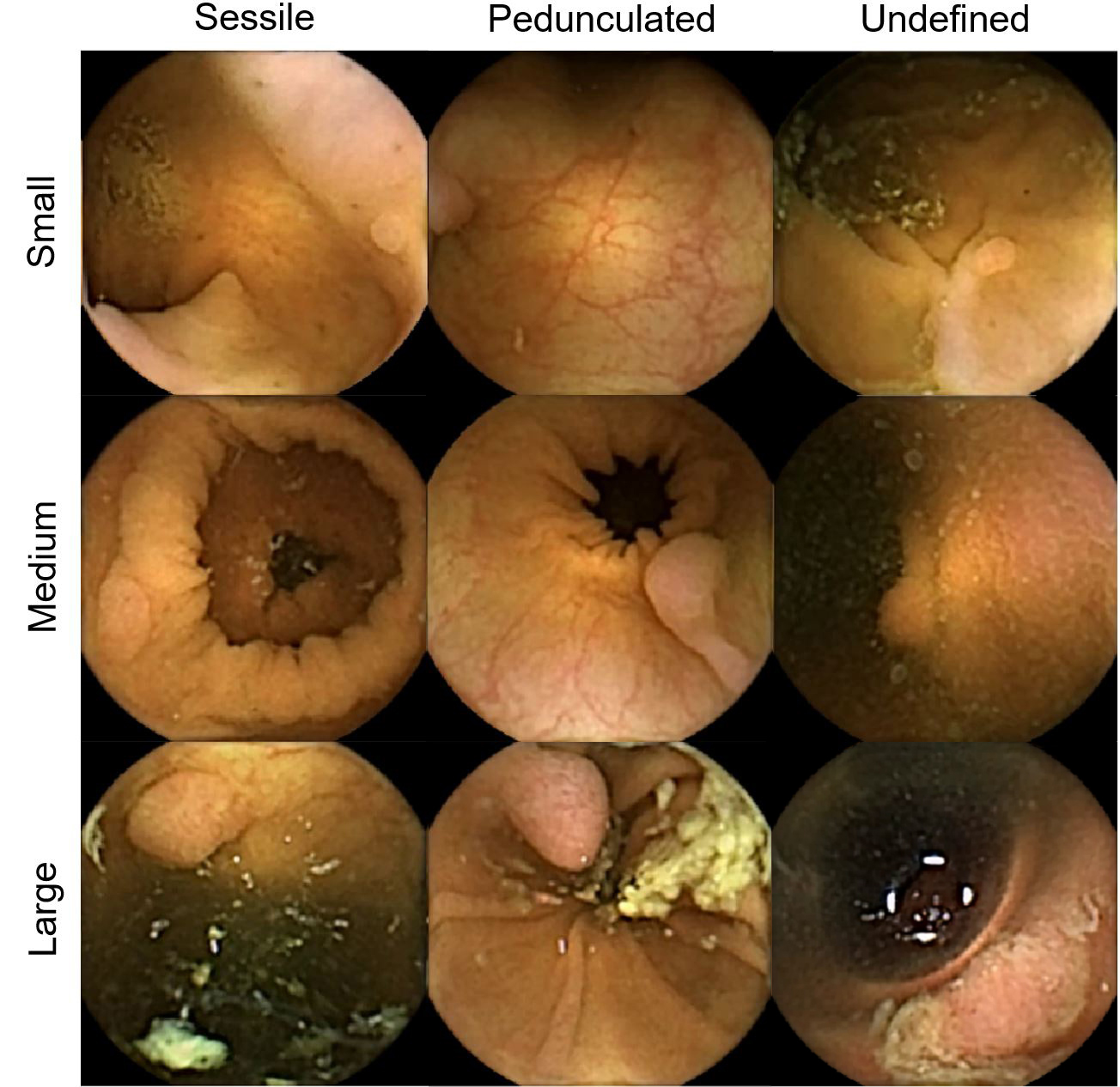}
    \caption{Polyp samples of different sizes and morphologies. Columns represents the different morphologies while raw represents the different sizes of the polyps.}
    \label{fig:polyps_size_morphology}
\end{figure}

Figure \ref{fig:polyps_size_morphology} shows 9 polyp samples of different sizes and morphologies. 
 
All the images have $256 \times 256$ resolution and the time stamp and device information were  removed.

\subsection{Architecture and Evaluation Details}

In all of the experiments, a pretrained model with ImageNet dataset is used to alleviate the problem of data scarcity. Moreover, in order to enlarge the amount of available images, data augmentation for training is performed by applying rotations of $0, 90, 180, 270$ degrees, horizontal and vertical flips and changes in the brightness of the images with a random probability.

Networks were optimized using Stochastic Gradient Descent with a fixed learning rate of $10^{-3}$ during 50 epochs. The hyper-parameter margin of the TL was fixed to $0.2$.  The batch size was fixed to 64, and the proportion of positive and negative images was set to 1/5 and 4/5 respectively. In order to not create a bias on the large videos, negative images were obtained using stratified random sampling from those procedures in the training set. Since the dataset is highly imbalanced, an epoch is considered once the entire set of positive images is passed forward and backward through the neural network.

To assess performance, results are reported following the 5-fold cross-validation strategy. It is important to remark that the stratified partitions have been done not by individual frames but by patients, thus images from the same patient do not belong to the same partition of the validation set.

The implementation of the methods has been done using TensorFlow and executed on a machine with an NVIDIA GeForce RTX 2080 TI. Training the network for 50 epoch takes about one hour, and the processing time per image in a forward pass is approximately $2.7ms$. Taking into consideration that the mean number of frames per procedure in our database is $15,183$, the mean processing time per video is $40.89s$. 

\subsection{Quantitative Results}

\begin{figure}
    \centering
    \includegraphics[width=.5\textwidth]{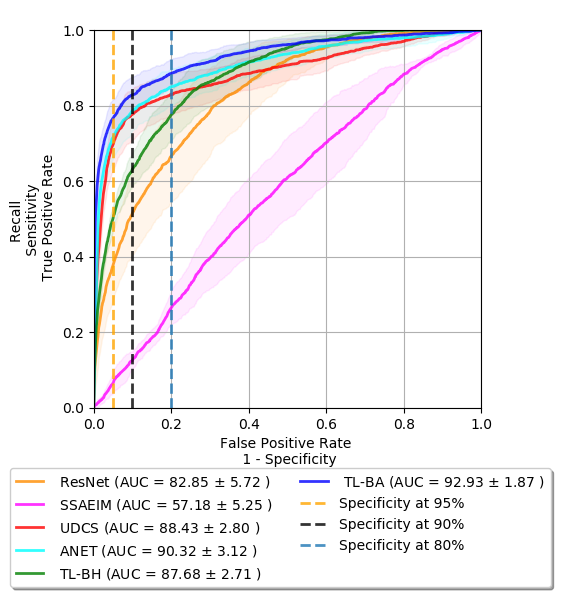}
    \caption{ROC Curve of the three models. Each vertical line represents a specificity value that indicates the percentage of true negative images predicted in the video, and the percentage of polyps that the system is expected to detect.}
    \label{fig:roc_cuts}
\end{figure}

In the first experiment, we aim to compare the performance of each one of the methodologies explained previously: \textit{ResNet}, \textit{$TL_{BH}$} and \textit{$TL_{BA}$}. 
As shown in Table \ref{results-summary}, our methodology \textit{$TL_{BA}$} has outperformed the obtained results by the standard optimization methodology of \textit{ResNet} and  \textit{$TL_{BH}$}. Contrary to what usually happens, batch all sampling strategy exceeds batch hard strategy. To our understanding, this result is due to the complexity of the generated triplets. Images from the same class are not visually similar. On one hand, all (positive and negative) images presents a high visual variability due to the free movement of the camera, the different parts of the gut or because of intestinal content, as food in digestion or bile, and on the other hand, polyps can be found in different stages presenting different sizes and morphologies. The obtained AUC value of our methodology has a $12\%$ increase respect to \textit{ResNet} and a $5\%$ compared to \textit{$TL_{BH}$}, achieving $92.94\pm 1.87\%$. The system enhancement is also reflected in the obtained sensitivity scores, which increased between $25$ and $40$ points compared to the other models. This fact shows that, given the same percentage of frames reviewed by the experts, the system finds more pathological frames.
Figure \ref{fig:roc_cuts} shows the ROC curves of the three studied models. 
On the left side of the curves, the \textit{$TL_{BA}$} model obtains a higher recall value than the other methods considering the same specificity. This difference means that \textit{$TL_{BA}$} detects more frames containing polyps, while at the right side of the curve the three  systems, \textit{$TL_{BA}$}, \textit{$TL_{BH}$} and \textit{ResNet}, work similarly.
It is remarkable to notice that in the \textit{$TL_{BA}$} experiment, the low standard deviation values indicate that the model is more robust than the others.

\begin{table*}[ht]
    \centering
    \caption{Performance comparison of the methods: ResNet, $TL_{BA}$ and $TL_{BH}$. Each method has been evaluated with a 5-fold validation in the classification task.}
    \label{results-summary}
    \resizebox{\linewidth}{!}{
    \begin{tabular}{@{}c|ccc||ccccc@{}}
    \toprule
     \textbf{Parameter} & \textbf{Accuracy} & \textbf{Sensitivity} & \textbf{Specificity} & \textbf{AUC} & \multicolumn{3}{c}{\textbf{Sensitivity} (\%)} \\ 
     
     \textbf{Optimization} & (\%) & (\%) & (\%) & (\%) & \textbf{Spec. at 95\%} & \textbf{Spec. at 90\%} & \textbf{Spec. at 80\%} & \\ 
     \midrule
    \textbf{ResNet} 
    & $ 97.85 \pm 0.24 $ &$ 26.01 \pm 9.78 $ &$ 97.97 \pm 0.27 $ &$ 82.85 \pm 5.72 $ &$ 37.75 \pm 9.12 $ &$ 51.49 \pm 11.09 $ &$ 66.71 \pm 12.15 $ \\ 
    
    \textbf{SSAEIM} &  $ 59.85 \pm 48.75 $ &$ 40.11 \pm 48.90 $ &$ 59.91 \pm 48.92 $ &$ 57.76 \pm 5.83 $ &$ 6.98 \pm 2.99 $ &$ 13.29 \pm 3.56 $ &$ 27.82 \pm 5.96 $  \\ 
    
    \textbf{UDCS} & $ 94.41 \pm 1.53 $ &$ 71.51 \pm 7.80 $ &$ 94.45 \pm 1.54 $ &$ 88.64 \pm 2.87 $ &$ 70.44 \pm 6.53 $ &$ 78.22 \pm 6.46 $ &$ 83.31 \pm 5.18 $  \\ 
    
    \textbf{ANET} 
    &   $ 96.96 \pm 0.53 $ &$ \mathbf{65.07 \pm 7.58} $ &$ 97.02 \pm 0.54 $ &$ 90.44 \pm 3.23 $ &$ 72.02 \pm 6.03 $ &$ 78.92 \pm 5.59 $ &$ 85.23 \pm 4.98 $  \\ 
    
    \midrule
    
    \textbf{$TL_{BH}$}  & $ \mathbf{99.83 \pm 0.05} $ &$ 0.00 \pm 0.00 $ &$ \mathbf{100.00 \pm 0.00} $ &$ 87.68 \pm 2.71 $ &$ 50.15 \pm 3.21 $ &$ 63.19 \pm 4.48 $ &$ 77.52 \pm 6.70 $ \\ 
    
    \textbf{$TL_{BA}$} 
    & $ \mathbf{99.43 \pm 0.12} $ &$ 51.15 \pm 7.62 $ &$ 99.51 \pm 0.12 $ &$ \mathbf{92.94 \pm 1.87} $ &$ \mathbf{76.68 \pm 4.93} $ &$ \mathbf{82.86 \pm 4.78} $ &$ \mathbf{88.53 \pm 3.76} $ \\

     
    
    \bottomrule
    \end{tabular}
    }
\end{table*}

The proposed method is also compared against some of the most recent polyp recognition systems: 1) \textit{SSAEIM} from \cite{11_deep_polyp_recognition}, 2) \textit{UDCS} from \cite{Yuan2019DenselyCN} and 3) \textit{ANET} from \cite{Guo2019TripleANet}.
All these methods have been implemented, trained and evaluated using the same dataset and evaluation methodology. The full details of the results are shown in Table \ref{results-summary} and Figure \ref{fig:roc_cuts}. As it can be observed, the proposed \textit{$TL_{BA}$} model demonstrates a significant improvement over these state-of-the-art systems.
The system shows an increment of at least the $2\%$ in the AUC value and an increase of around $4\%$ in the different sensitivity values.
On one hand, we observed that \textit{SSAEIM} method is not able to handle this imbalanced and complex database. From our point of view, as a consequence of the complexity of the problem, the dense layers of the autoencoders were not able to find a proper representation of the images.
On the other hand, and although \textit{UDCS} and \textit{ANET} show that good results, the obtained AUC score and the sensitivity values do not reach the results obtained with our method.
This result reflects the capacity of our method to detect polyps, validating the efficiency of TL in this application.

In the next experiment, see Table \ref{results-embeddings},  we contrast the performance of $TL_{BA}$ against the same methodology but adding a new extra layer, that acts as the embedding layer. This experiment is done since when deep metric losses are used, it is common to add an extra dense layer between the extracted features and the classification layer. This layer introduces more versatility in data representation while it compresses the information in the embedding. As shown in Table \ref{results-embeddings}, the embedding sizes used in these experiments are: $128$, $256$, $512$ and $1,024$.
Despite the fact that the new networks have more parameters, none of them exceeds the previous $TL_{BA}$ results in the AUC score.
The obtained AUC value and sensitivity values show a correlation between the embedding size and the obtained scores.
$TL_{BA}$ model with an embedding size of 2,048 have exceed the other models because the variation on the embedding size allows the network to have a better representation to detect the polyps.

\begin{table*}[ht]
    \centering
    \caption{Performance of the methods: $TL_{BA}$ and different versions of the same adding an extra dense layer and changing the embedding size. Each method has been evaluated with a 5-fold validation in the classification task.}
    \label{results-embeddings}
    \resizebox{\linewidth}{!}{
    \begin{tabular}{@{}c|ccc||ccccc@{}}
    \toprule
     & \textbf{Accuracy} & \textbf{Sensitivity} & \textbf{Specificity} & \textbf{AUC} & \multicolumn{3}{c}{\textbf{Sensitivity} (\%)} \\ 
     
    \textbf{Embedding} & (\%) & (\%) & (\%) & (\%) & \textbf{Spec. at 95\%} & \textbf{Spec. at 90\%} & \textbf{Spec. at 80\%} \\ 
    \midrule
    \textbf{$TL_{BA}$} 
    & $ \mathbf{99.43 \pm 0.12} $ &$ 51.15 \pm 7.62 $ &$ \mathbf{99.51 \pm 0.12} $ &$ \mathbf{92.94 \pm 1.87} $ &$ \mathbf{76.68 \pm 4.93} $ &$ \mathbf{82.86 \pm 4.78} $ &$ \mathbf{88.53 \pm 3.76} $ \\

     \textbf{128} & $ 59.99 \pm 7.48 $ &$ \mathbf{84.67 \pm 6.17} $ &$ 59.95 \pm 7.51 $ &$ 83.05 \pm 2.78 $ &$ 42.23 \pm 5.04 $ &$ 57.38 \pm 5.54 $ &$ 72.54 \pm 4.85 $ \\ 
    
     \textbf{256} & $ 68.12 \pm 12.92 $ &$ 79.10 \pm 14.60 $ &$ 68.09 \pm 12.97 $ &$ 83.35 \pm 5.27 $ &$ 45.09 \pm 10.34 $ &$ 59.9 \pm 11.64 $ &$ 73.35 \pm 10.11 $ \\ 
    
     \textbf{512} & $ 86.29 \pm 2.56 $ &$ 70.50 \pm 9.02 $ &$ 86.31 \pm 2.57 $ &$ 86.58 \pm 3.91 $ &$ 49.91 \pm 8.56 $ &$ 64.66 \pm 6.43 $ &$ 78.13 \pm 6.04 $ \\ 
    
     \textbf{1024} & $ 89.69 \pm 2.50 $ &$ 68.28 \pm 9.44 $ &$ 89.72 \pm 2.51 $ &$ 86.67 \pm 3.25 $ &$ 55.47 \pm 3.86 $ &$ 67.97 \pm 5.79 $ &$ 79.89 \pm 5.29 $ \\ 
    
     \bottomrule
    \end{tabular}
    }
\end{table*}


The margin hyper-parameter of the TL has been set until now at $0.2$ as it is set in other works like \cite{triplet_loss} or \cite{defense_triplet}.
As the domain of the problem is different from previous applications of the TL method, our fourth experiment evaluates the behaviour of the system with the following margins: $0.1$, $0.5$ and $1.0$.
As shown in the obtained results summarized in Table \ref{results-margins}, any of these margins outperforms all the metrics. 
Margins $0.5$ and $1.0$ obtain standard deviation values which are higher than the small margins, indicating that the model is less robust.
However, the margin that achieves the best results on almost all the reported metrics is $0.2$. 

\begin{table*}[ht]
    \centering
    \caption{Performance of $TL_{BA}$ method changing the margin parameter. 
     Each network has been evaluated with a 5-fold validation in the classification task.}
    \label{results-margins}
    \resizebox{\linewidth}{!}{
    \begin{tabular}{@{}c|ccc||ccccc@{}}
    \toprule
      & \textbf{Accuracy} & \textbf{Sensitivity} & \textbf{Specificity} & \textbf{AUC} & \multicolumn{3}{c}{\textbf{Sensitivity} (\%)}\\ 
     
     \textbf{Margin} & (\%) & (\%) & (\%) & (\%) & \textbf{Spec. at 95\%} & \textbf{Spec. at 90\%} & \textbf{Spec. at 80\%} \\ 
     \midrule
    
    \textbf{0.1} & $ 87.98 \pm 2.01 $ &$ 66.85 \pm 7.75 $ &$ 88.01 \pm 2.01 $ &$ 86.53 \pm 2.96 $ &$ 47.35 \pm 10.22 $ &$ 63.35 \pm 7.66 $ &$ 78.69 \pm 5.60 $  \\

    \textbf{0.2 ($TL_{BA}$)} & $ \mathbf{99.43 \pm 0.12} $ &$ 51.15 \pm 7.62 $ &$ \mathbf{99.51 \pm 0.12} $ &$ \mathbf{92.94 \pm 1.87} $ &$ \mathbf{76.68 \pm 4.93} $ &$ \mathbf{82.86 \pm 4.78} $ &$ \mathbf{88.53 \pm 3.76} $ \\
    
    \textbf{0.5} & $ 90.89 \pm 2.07 $ &$ \mathbf{68.79 \pm 8.66} $ &$ 90.93 \pm 2.07 $ &$ 87.74 \pm 3.04 $ &$ 56.07 \pm 9.18 $ &$ 70.96 \pm 7.44 $ &$ 83.87 \pm 4.61 $ \\

     \textbf{1.0} & $ 91.67 \pm 1.00 $ &$ 66.30 \pm 8.74 $ &$ 91.71 \pm 1.02 $ &$ 86.80 \pm 3.23 $ &$ 56.84 \pm 8.68 $ &$ 70.65 \pm 7.38 $ &$ 81.02 \pm 6.06 $ \\
    
     \bottomrule
    \end{tabular}
    }
\end{table*}

The comparison of models is shown in Tables \ref{results-summary}, \ref{results-embeddings} and \ref{results-margins} demonstrating that \textit{$TL_{BA}$} is the best computer-aided decision support system for polyp detection in terms of accuracy and sensitivity.

\begin{figure}
    \centering
    \includegraphics[width=.43\textwidth]{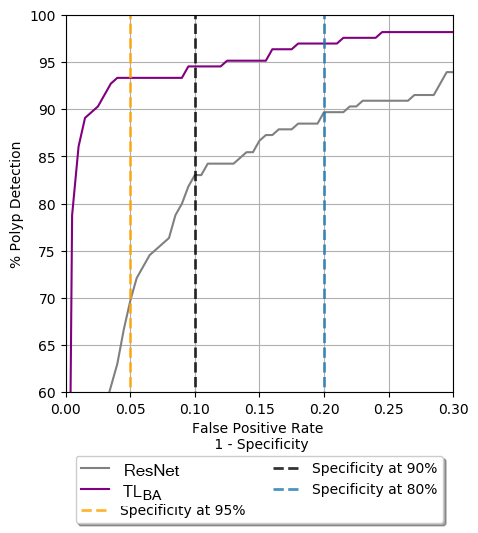}
    \caption{ Percentage of polyps detected.}
    \label{fig:polyp_detection}
\end{figure}

From a medical point a view, the computer-aided system should help to detect polyps but not necessarily detect all the images where a given polyp is seen. 
For this reason, we analyzed the performance of our proposed system over polyps.
A global overview of the numerical results is summarized in Table \ref{tab:recall_specificity}, where each score represents the percentage of detected polyps in different scenarios of the entire dataset.
Each of them is computed with a different specificity value: $80\%$, $90\%$ and $95\%$.
The first row of the table contains the percentage of detected polyps, that grows when the specificity decreases.
Setting the specificity at $95\%$, the system only misses  $14$ polyps; if we decrease specificity to $90\%$ and $80\%$, the missed polyps are $11$ and $8$ respectively. 
A complete view of the curve is reported in Figure \ref{fig:polyp_detection}.

The second set of results in Table \ref{tab:recall_specificity} present the detection of the system according to polyp size.
When we consider small polyps  the amount of missed polyps is 7, 4 and 3 for the respectively reported values of specificity.  
In the case of medium-sized polyps,  using the specificity of $90$ or higher, the system is not able to detect $5$ polyps, but with lower specificity, the amount of detected polyps raises. 
In the case of larger polyps,  2 of them are lost for the two higher specificity values, however, with a slight decrease in specificity, the polyps are detected.

Finally, the lasts rows of Table \ref{tab:recall_specificity} show the detection rated based on polyp morphology: sessile, pedunculated or undefined.
As it has been reported previously, most polyps are labeled as sessile, obtaining high detection scores despite the misplacement of $7$. 
Pedunculated polyps are relatively rare, and the system detected all except one at the three sensitivity values.

\begin{table}[H]
\centering
\caption{Detection vs. Specificity with model $TL_{BA}$}
\label{tab:recall_specificity}
\resizebox{!}{!}{
    \begin{tabular}{@{}c|ccc@{}}
    \toprule
    \textit{\textbf{$\%$ detection}} & \textit{\textbf{Specificity@95}} & \textit{\textbf{Specificity@90}} & \textit{\textbf{Specificity@80}}\\ 
    \midrule
    
    \textit{\textbf{Polyps}} & $91.14\%$ & $93.04\%$ &  $94.94\%$\\ \midrule

    \textit{\textbf{Small Polyps}}& $91.86\%$ & $95.35\%$ & $96.51\%$\\
    \textit{\textbf{Medium Polyps}}& $90.38\%$ & $90.38\%$ & $92.31\%$ \\
    \textit{\textbf{Large Polyps}}& $90.00\%$ & $90.00\%$ & $95.00\%$ \\
    \midrule

    \textit{\textbf{Sessile Polyps}}& $93.07\% $ & $95.05\%$ & $96.04\%$ \\
    \textit{\textbf{Pedunculated Polyps}}& $90.91\% $ & $90.91\%$ & $100.00\%$ \\
    \textit{\textbf{Undefined Morphology}}& $86.96\% $ & $89.13\%$ & $91.30\%$ \\
    \bottomrule
    \end{tabular}
}
\end{table}

\subsection{Qualitative Results and Polyp Localization}

\begin{figure*}[]
    \centering
    \includegraphics[width=1\textwidth]{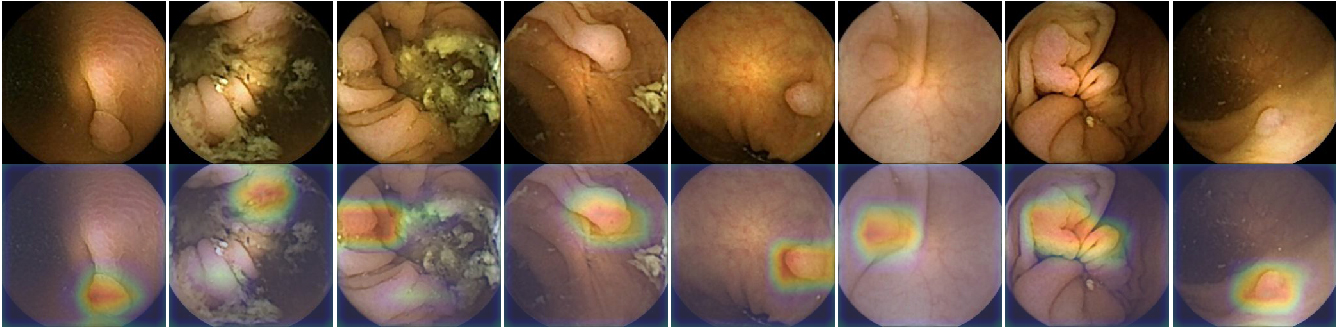}
    \caption{The first row contains eight examples of TP of our proposed method, with polyps of different morphology and size.
    The second row incorporates the CAM representation that locates each one of the polyps over the original image. }
    \label{fig:true_positive_sample}
\end{figure*}

CAM visualization was applied to the output of the network.
This method generates a heat map, where the red tones show the regions of the image that obtain a high response from the filters.
Figure \ref{fig:true_positive_sample} shows in the first row eight polyps frames where the different morphology and size of the polyps may be observed.
In the second row, the CAM visualization method highlights the location where the system focused to predict that there was a polyp.

\begin{figure*}[]
    \centering
    \includegraphics[width=1\textwidth]{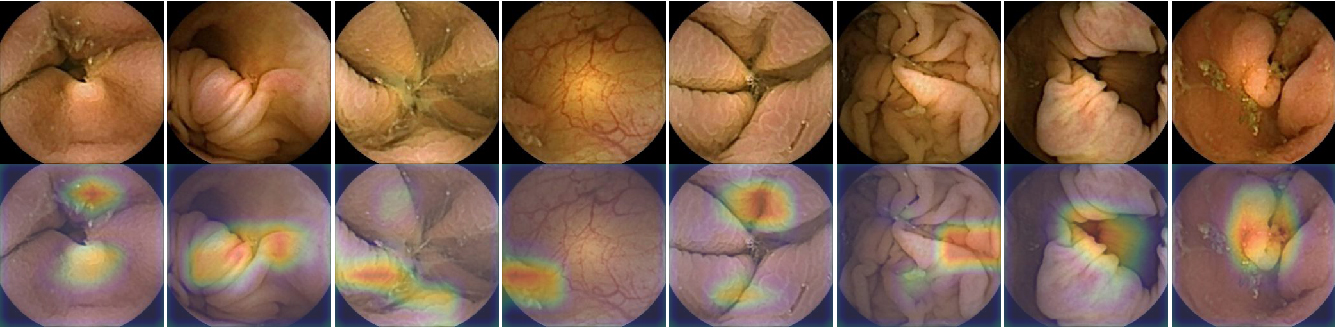}
    \caption{The first row contains eight examples of FP of our proposed method, where the system has detected polyps. In some images abnormal tissue can be seen, some mucous membrane or reddish zone, that are features related to polyps.
    The second row shows the CAM representation that locates where these features are located.}
    \label{fig:false_positive_sample}
\end{figure*}

Figure \ref{fig:false_positive_sample} shows eight images without polyps where the system has erroneously detected a polyp.
In these samples, some of the regions highlighted by the network contain features of polyps such as growths of tissue, mucous membranes or areas with reddish colour from the wall, that might indicate the existence of it.

\begin{figure*}[]
    \centering
    \includegraphics[width=1\textwidth]{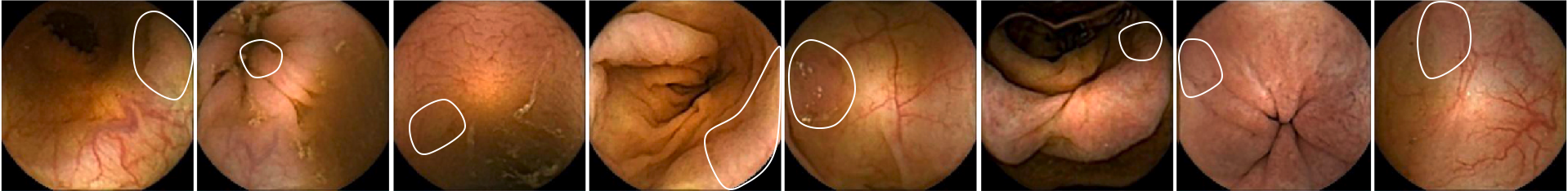}
    \caption{The images correspond to eight examples of FN of our proposed method, where a polyp is in the frame, but the system couldn't detect it. To help the reader to find the polyps in the images, the outline of the polyp has been drawn in white color.}
    \label{fig:false_negative_sample}
\end{figure*}

Figure \ref{fig:false_negative_sample} shows eight polyp images where the system has not obtained enough features to predict the frame as polyp.
Each image shows a boundary with the location of the polyps. 
These difficult cases are complex to detect in single images by the system. The evaluation of a whole sequence of images where the polyp is seen facilitates detection by the human eye. Due to the complexity of polyp detection, sometimes is easier for humans to detect them through the sequence.

\begin{figure*}[]
    \centering
    \includegraphics[width=1\textwidth]{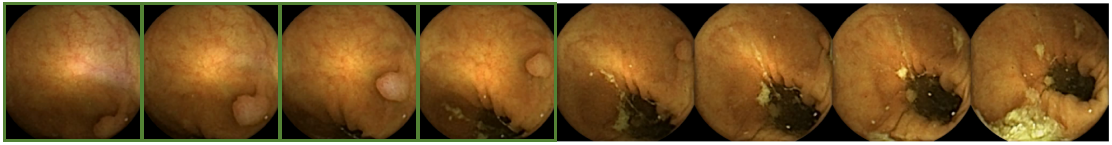}
    \caption{Polyp sequence where the green squares denote the presence of polyps detected by the system. In this sequence, there are two frames where the polyp is not detected, despite this, the support system has found the polyp in the previous frames, allowing the doctor to diagnose the patient.}
    \label{fig:detection_sequence}
\end{figure*}

Figure \ref{fig:detection_sequence} shows the second sequence of images in Figure \ref{fig:sample_seq} with the output of the system represented by adding a green square around the frames where the system has detected a polyp.
Although in this example the system missed two frames where the polyp is present, the detection in four frames is sufficient for the physician to establish the diagnosis.

\subsection{Effect of imbalance datasets over models}

Healthcare datasets commonly suffer from imbalanced data, a feature that frequently affects the performance of classical deep learning approaches. In the following experiment, we evaluate how both, TL and the original ResNet networks behave with respect to different degrees of imbalanced data. The imbalance degree of a dataset is defined as the number of negative images per each positive one. 
To conduct a meaningful comparison, both models are trained and evaluated on six different imbalance degrees (1, 10, 25, 50, 100). Furthermore, each imbalance degree is repeated ten times using different sampled data. 

\begin{figure}
    \centering
    \includegraphics[width=.5\textwidth]{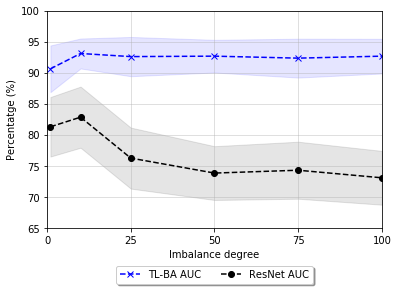}
    \caption{AUC values of ResNet and our methodology trained with difference imbalance degrees. Each row represent the mean of 10 executions of a 5-fold validation in the classification task.}
    \label{fig:imb_degree_both}
\end{figure}

As it can be seen in Figure \ref{fig:imb_degree_both}, the experiment shows the robustness of the proposed methodology to highly imbalanced data. The performance of the method increases when the degree of imbalance increased from 1 to 10 and then stabilizes. On the other hand, it can be observed that the original ResNet network suffers with highly imbalanced data. The network achieves the best result with an imbalance degree of 10 and then it starts to deteriorate as the imbalance degree increases.

\section{Conclusion}

The methodology proposed in this study improves automatic polyp detection in WCE images and additionally enables localization of the polyp in each image.
The reported experiments demonstrate that the TL method improves feature extraction outperforming previous results and that the limited and imbalanced data availability may be alleviated with the appropriate losses.
Furthermore, the qualitative output of the system may increase trust in the prediction.

Future research will focus on the detection of other intestinal pathologies to develop a complete computer-aided detection system for WCE videos.

\section*{Acknowledgment}

The authors would like to thank the team from CorporateHealth International ApS for their feedback and economic support and NVIDIA for their GPU donations. This work has been also supported by MINECO Grant RTI2018-095232-B-C21 and SGR 1742.

\bibliographystyle{unsrt}  
\bibliography{references}  


\end{document}